\documentclass{IEEEtran}
\usepackage{cite}
\usepackage{amsmath,amssymb,amsfonts}
\usepackage{graphicx}
\usepackage{textcomp,nicefrac}
\def\BibTeX{{\rm B\kern-.05em{\sc i\kern-.025em b}\kern-.08em
T\kern-.1667em\lower.7ex\hbox{E}\kern-.125emX}}
\begin{document}
\title{ A data processing system for balloon-borne telescopes}

\author{Valentina Scotti~\IEEEmembership{Member,~IEEE},
	    Giuseppe Osteria,
	    and Francesco Perfetto
\thanks{Manuscript received October 30, 2020. This work was partially supported by the Italian Space Agency through the ASI INFN agreement n. 2017-8-H.0.}
\thanks{V. Scotti is with the Department of Physics of University of Naples Federico II and INFN Naples, Italy (email: scottiv@na.infn.it)}
\thanks{G. Osteria is with INFN Naples, Italy}
\thanks{F. Perfetto is with INFN Naples, Italy}
}


\maketitle

\begin{abstract}
The JEM-EUSO Collaboration aims at studying Ultra High Energy Cosmic Rays (UHECR) from space. To reach this goal, a series of pathfinder missions has been developed to prove the observation principle and to raise the technological readiness level of the instrument. Among these, the EUSO-SPB2 (Extreme Universe Space Observatory on a Super Pressure Balloon, mission two) foresees the launch of two telescopes on an ultra-long duration balloon. One is a fluorescence telescope designed to detect UHECR via the UV fluorescence emission of the showers in the atmosphere. The other one measures direct Cherenkov light emission from lower energy cosmic rays and other optical backgrounds for cosmogenic tau neutrino detection.

In this paper, we describe the data processing system which has been designed to perform data management and instrument control for the two telescopes. It is a complex which controls front-end electronics, tags events with arrival time and payload position through a GPS system, provides signals for time synchronization of the event and measures live and dead time of the telescope. In addition, the data processing system manages mass memory for data storage, performs housekeeping monitor, and controls power on and power off sequences.

The target flight duration for the NASA super pressure program is 100 days, consequently, the requirements on the electronics and the data handling are quite severe. The system operates at high altitude in unpressurised environment, which introduces a technological challenge for heat dissipation.
\end{abstract}


\section{Introduction}
\label{sec:introduction}
\IEEEPARstart{T}{he} EUSO-SPB2 is a precursor mission for future space observatories for multi-messenger astrophysics. 

The innovative approach of EUSO-SPB2 is to use the Earth as a giant particle detector, studying fluorescence and Cherenkov light from the edge of the space. 

The main goal of this mission is to detect fluorescence light generated in the Earth's atmosphere by Ultra High Energy Cosmic Rays (UHECR) from above, to confirm the expectations from ground observations \cite{auger,ta}. 

In addition, this mission will explore new areas such as detecting Cherenkov light emission from air showers and measuring the background of up-going tau decays from cosmogenic neutrinos \cite{poemma}.

This mission is part of the JEM-EUSO science program. JEM-EUSO is a collaboration of 350 researchers from 16 countries, whose scientific case has been positively evaluated by several space agencies with funding ongoing in all the participating countries. 

The aim of JEM-EUSO is to observe from space UV light in the terrestrial atmosphere to study Extreme Energy Cosmic Rays (EECR): above $5\times10^{19}$eV. The main challenge for the study of EECR is the extremely low flux of cosmic rays at the highest energies.

The scientific program of JEM-EUSO includes a series of pathfinder experiments using fluorescence detectors to make a proof-of-principle of the EECR observation from space and to raise the technological level of the instrumentation to be employed in a space mission.

Among these, the EUSO-SPB2 is build upon the EUSO-SPB1 \cite{euso-spb1} experience to pave the way towards the Probe of Extreme Energy Multi-Messenger Astrophysics (POEMMA) space mission \cite{poemma}. POEMMA is designed to discover the nature and origin of EECR and to discover astrophysical neutrino emission. 

The launch of the EUSO-SPB2 for an ultra-long duration balloon flight is foreseen in 2022 from Wanaka in New Zealand. The target flight duration for the NASA super pressure balloon program is 100 days at a floating altitude of 33 km.

A description of the mission is given in \cite{euso-spb2}, in this paper, we present the concept of the Data Processor system designed to manage the main instruments and all the devices on board while respecting the severe constraints imposed by a balloon-borne experiment.

\section{The EUSO-SPB2 telescopes}

\begin{figure}[bt]
\centerline{\includegraphics[width=0.9\linewidth]{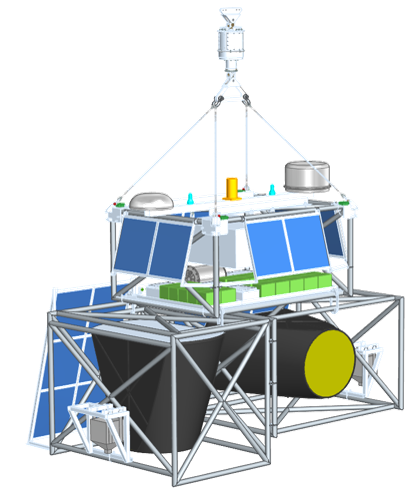}}
\caption{A pictorial view of the gondola hosting the two telescopes and all the ancillary devices of the EUSO-SPB2 mission. }
\label{fig:gondola}
\end{figure}

EUSO-SPB2 will host onboard two independent telescopes, a fluorescence detector (FT) and a Cherenkov detector (CT), in order to be a scientific and technical sub-orbital altitude precursor for the POEMMA. A complete description of the two instruments can be found respectively in \cite{fluorescence} and \cite{cherenkov}, here we will briefly report only on the details relevant to the functioning of the Data Processing system. Fig. \ref{fig:gondola} shows a drawing of the gondola structure.

In table \ref{tab:tab1} the main characteristics of the two telescopes are summarized.
\begin{table}
\caption{Main parameters of the EUSO-SPB2 telescopes. The two telescopes share the same mechanical structure, mirror, and a 1-meter wide PMMA corrector plate.}
\label{table}
\setlength{\tabcolsep}{3pt}
\begin{tabular}{|c|c|c|}
\hline
Fluorescence &  & Cherenkov \\
\hline
 & Energy threshold (eV) &  \\
 & Sensor type & \\
  300$\div$420 nm, BG3 filter & Wavelenght sensitivity & 400$\div$800 nm, no filter\\
 & Time unit& \\
 & Pointing (zenith angle) & \\
 & FoV (instrumented) & \\
 & Number of pixels & \\
 & Optics (modified Schmidt) & \\
 & Payload mass (kg)& \\
 \hline
\end{tabular}
\label{tab:tab1}
\end{table}

The fluorescence detector is build on the experience of EUSO-SPB1 to detect ultra high energy cosmic rays via the UV fluorescence emission of the extensive air showers generated by UHECR in the atmosphere. The main difference with respect to EUSO-SPB1 is that in EUSO-SPB2 the light is focused on the camera through a Schmidt optical system, instead of Fresnel lenses. In addition, the focal surface of the FT will be composed of 3 Photo Detector Modules (PDM), each composed of 36 MAPMT Hamamatsu R11265, for a total of 6912 pixels, compared to 2304 of EUSO-SPB1. The Field of View of each PDM is $12^{\circ}\times12^{\circ}$. The signal from MaPMT are with are readout by a specific ASIC followed by multiple trigger levels to filter out noise and identify events of interest. This system has single photoelectron counting capability and a time unit of $1 \mu s$. The FT points to nadir and the expected trigger rate is around 0.2/hour for a trigger threshold of $10^{18}$eV. 

The CT points toward the limb to measure direct Cherenkov light emission from lower energy cosmic rays and the optical backgrounds for cosmogenic tau neutrino detection in the UV/VIS range. For the CT the mirror segments will be aligned in a bifocal configuration which allows to distinguish between a charged particle hitting camera which makes 1 spot, and a light pulse from far-field outside the telescope makes 2 spots. The CT focal surface will be composed by 32 Hamamatsu S14521 (4 × 4 pixel array) SiPMT, for a total of 512 pixels. The readout is based on a 100 MS/s ASIC for General Electronics for TPC‘s (AGET) switch capacitor ring sampler.

Each telescope is completed by several ancillary devices:
\begin{itemize}
    \item Aperture shutter to prevent sun-melt and avoid dust entering the camera. The shutter is light tight and makes the inside of the telescope a dark box.
    \item Health LED Systems to check camera health and response.
    \item Airglow MONitor (AMON) for an independent measure of airglow and night sky radiation backgrounds.
    \item Light sensor, called EMON, located inside the telescopes to verify darkness condition during operation and testing.
    \item A dual-band InfraRed camera to monitor the presence of clouds and to measure cloud-top temperatures only for the fluorescence telescope.
\end{itemize}

\section{The data processing system}

The focal plane detectors and the front-end electronics of the two telescopes are different since the Cherenkov and fluorescence light have different wavelengths and time development. However, from the point of view of the control, configuration, monitoring, data management and working mode, the two instruments are very similar. For this reason, the architecture of the Data Processing system designed for the two telescopes is an evolution of the one developed for the fluorescence telescope of the EUSO-SPB1 \cite{dp}.

\begin{figure*}[bt]
\centerline{\includegraphics[width=\textwidth]{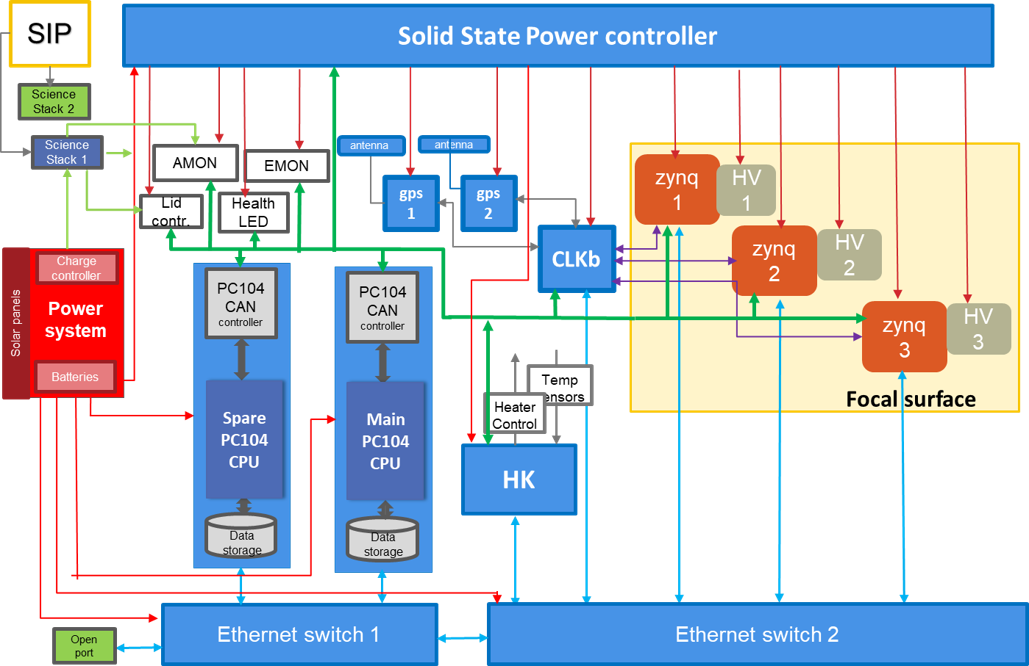}}
\caption{Block diagram of the Data Processing system nd its connections with the rest of the instrument for the fluorescence telescope of the EUSO-SPB2. The one for the Cherenkov detector is quite similar. }
\label{fig:blockdiag}
\end{figure*}

The block diagram of the DP of the fluorescence telescope and its connections with the rest of the instrument is shown in Fig.~\ref{fig:blockdiag}. 
The one for the Cherenkov detector is quite similar. In the following paragraph we will describe the Data Processing system of the FT.

\subsection{Requirements}

The EUSO-SPB2 telescopes will collect data during moonless nights to obtain the
best contrast for the signatures of scientific interest. The data processing system (DP) links each telescope with the Gondola system. The DP is a complex system which includes most of the digital electronics of the instrument which allows to control, configure, monitor and operate each telescope during the commissioning phase, the test campaigns and the flight. 

Besides being the interface with the flight computer, the data processing system performs the following tasks:
\begin{itemize}
        \item Main interface with Flight Computer (SIP) telemetry system
        \item Data selection/compression and transmission to Flight Computer (SIP)
        \item Power ON/OFF the whole instrument
        \item Define Telescope operation mode (Day, Night, D-N-D transitions)
        \item Configure the Front End electronics
        \item Start/Stop of the data acquisition and calibration procedures
        \item Tag events with GPS time and GPS position
        \item Synchronization of the data acquisition
        \item Manage trigger signals (L1, external, GPS, by CPU command etc) 
        \item Manages mass memory for data storage
        \item Monitor/Control/DAQ of some Ancillary Devices
        \item Monitor voltages, current and temperatures (LVPSs, boards, FPGAs)
\end{itemize}

The main challenges in designing and building the system are due to the physical characteristics of the signal and to the harsh environment in which the DP has to operate.

At the Peak energy sensitivity for the FT (around 6 EeV) the expected event rate is around 1 event every 13 hours. For an 80 day flight with 15\% duty cycle the total number of events is around 22 events, but the trigger rate of the FT is estimated to be between 5 and 10 Hz, so it’s necessary to distinguish a few cosmic rays events among the background. 

Furthermore, since the target flight duration for the NASA super pressure program is 100 days, the requirements on the electronics and the data handling are quite severe. Finally, The system operates at high altitude in unpressurised environment, which introduces a technological challenge for heat dissipation.

In addition to power and mass budget restrictions on the balloon payload, there is also a limited telemetry budget of 1GB per day. For this reason, there is the need to prioritize data for downloading on board, at the same time with the want to only record and transfer high quality events. Everything must be done on board with minimal intervention from the ground.

\section{The subsystems}

The block diagram (Fig. \ref{fig:blockdiag}) shows that the data processing system is composed of several subsystems. The whole system is capable to acquire 7000 channels without exceeding mass and power budget. ----------

\begin{figure}[bt]
\centerline{\includegraphics[width=\linewidth]{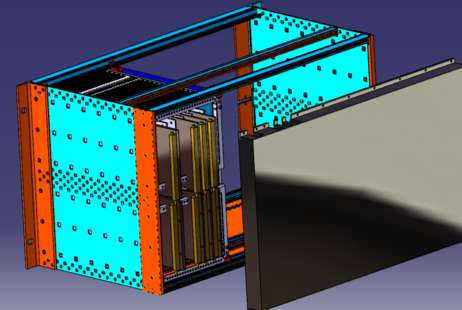}}
\caption{The CAD of the DP box, designed to host all the Data Processing subsystems and equipped with cooling plate. }
\label{fig:structure}
\end{figure}

Different tasks are performed by different subsystems. All the subsystems, together with the low voltage power supplies modules, are hosted in a DP box (customized Eurocard chassis) equipped with cooling plate to dissipate heat (Fig. \ref{fig:structure}).

\begin{itemize}
    \item CPU with Hot and Cold redundancy PCI/104 single board computer Core i7 3517UE + 5 SATA disks for data storage
    \item HK board board based on STM32 µController and TIBO EM2000 
    \item Clock board \cite{clock} based on Xilinx XC7Z SoC:  manages trigger and data synchronization and also acts as an interface with the GPS receivers
    \item GPS receivers Trimble BX992 
    \item Ethernet switches 
    \item Solid State Power Controller 
\end{itemize}

We choose to use mostly COTS devices because reliability in a hostile environment is one of the requirements. The electronic components are all selected to operate in extended temperature range.

For the most fundamental component, the CPU, we have hot and cold redundancy, hence the choice to use CAN and Ethernet protocol which guarantee us in case one CPU fails.

\section{The software}

The DP is the interface between the detector and the telemetry blocks and the end-users.Instrument operations and monitoring will be shared between collaborators in the US, Japan and Europe to provide, during local daytime, control and monitoring of the telescopes 24/7, similarly to the on-ground control that was done for EUSO-SPB1 \cite{software}.  

The CPU software can be divided into two categories:
\begin{itemize}
    \item the control software 
    \item the data handling software
\end{itemize}

The control software has to be flexible to be used in every phase of the commissioning and flight. It allows to control and monitor the detector, and to store all the data (science, housekeeping and log). The design enables full user control via telemetry. In fact, during the flight the status of the apparatus must be continuously checked to distinguish condition to start or stop the measurement.

The data handling software manages science data acquisition from the CLK and Zynq boards. The acquisition software initializes and configures the subsystems, monitors the connections with the CPU, and verifies the behavior of the subsystems. It can perform several types of acquisitions, such as recording of external triggered events or internal triggered events. Finally, it manages the prioritization folder structure for down-link to cope with the maximum amount of telemetry available each day (1 GB).

Science Stack data are available in ''quasi'' real time on a WEB page of Columbia Scientific Balloon Facilities and are used to monitor and to control the status of the telescope. In particular, before starting the power-on or power-off sequence, it is necessary to monitor:
\begin{itemize}
    \item Status of the power system
    \item Temperatures of critical point of the telescope (Batteries, mirror, focal surface, electronics, etc.)
    \item Light level of the EMON (Day/Night transition, Night/Day transition)
    \item Status of the aperture shutter (Open/Close)
\end{itemize}

\section{Schedule}

The launch of the EUSO-SPB2 is planned for 2022,  in order to reach this goal, the EUSO-SPB2 Collaboration need to hit several milestones.

CPUs, data storage, GPS receivers, Ethernet switches, Solid State Power Controller are under test in different laboratories. The first prototype of the Clock board has been manufactured and tested, the firmware is under development. The design of the HK board has been completed and the first prototype will be available before the end of 2020. The design of the DP housing and cooling mechanics are in progress and will be soon finalized and tested. The Collaboration is currently looking for facilities for thermal-vacuum tests of the whole DP. Extensive laboratory testing is planned to characterize and calibrated the telescopes prior to payload integration. 

At the end of the assembly, and after thermal-vacuum chamber test, both the telescopes will be transported to the Utah desert for a field test campaign. The aim of this campaign is to use lasers and other light sources to have an absolute calibration of the detector response. The field tests will include observation time to record Fluorescence and Cherenkov light due to cosmic rays. The results of the tests will be used also to fine-tune the trigger algorithm. 
.

\section{Conclusions}
EUSO-SPB2 is a 2-telescope instrument to study UHECR, Cherenkov, and tau neutrino backgrounds. The launch planned for 2022. The Data Processing system of the telescopes have been designed and the first prototypes are under test. The Collaboration is getting ready for thermal-vacuum test and instrument integration.

\section*{Acknowledgment}

We would like to thank the staff of Electronics and Detectors Service (SER) and  Design and Mechanics  Service of INFN Napoli for their strong and undivided support all along this project.

\end{document}